\documentclass[preprint,eqsecnum,aps,nofootinbib]{revtex4}
\usepackage{amsfonts,amsmath,amssymb,amsthm}
\usepackage{latexsym}
\usepackage{bbm,bm}
\usepackage{graphicx}

\newcommand{\ket}[1]{\lvert #1 \rangle}
\newcommand{\bra}[1]{\langle #1 \lvert}
\newcommand{\beq}{\begin{equation}}
\newcommand{\eeq}{\end{equation}}
\newcommand{\beqs}{\begin{eqnarray}}
\newcommand{\eeqs}{\end{eqnarray}}

\begin{document}

\title{Entanglement of Four-Qubit Rank-$2$ Mixed States}

\author{Eylee Jung$^{1}$ and DaeKil Park$^{1,2}$}

\affiliation{$^1$Department of Electronic Engineering, Kyungnam University, Changwon
                 631-701, Korea    \\
             $^2$Department of Physics, Kyungnam University, Changwon
                  631-701, Korea    
                      }

\begin{abstract}
It is known that there are three maximally entangled states $\ket{\Phi_1} = (\ket{0000} + \ket{1111}) / \sqrt{2}$,
$\ket{\Phi_2} = (\sqrt{2} \ket{1111} + \ket{1000} + \ket{0100} + \ket{0010} + \ket{0001}) / \sqrt{6}$, and 
$\ket{\Phi_3} = (\ket{1111} + \ket{1100} + \ket{0010} + \ket{0001}) / 2$ in four-qubit system. It is also 
known that there are three independent measures ${\cal F}^{(4)}_j \hspace{.2cm} (j=1,2,3)$
for true four-way quantum entanglement in the same system. In this paper we compute ${\cal F}^{(4)}_j$ 
and their corresponding linear monotones ${\cal G}^{(4)}_j$ for 
three rank-two mixed states $\rho_j = p \ket{\Phi_j}\bra{\Phi_j} + (1 - p) \ket{\mbox{W}_4} \bra{\mbox{W}_4}$, 
where $\ket{\mbox{W}_4} = (\ket{0111} + \ket{1011} + \ket{1101} + \ket{1110}) / 2$. We discuss 
the possible applications of our results briefly.
\end{abstract}

\maketitle

\section{Introduction}

Recently, much attention is being paid to quantum information theory (QIT) and quantum technology (QT)\cite{text}. Most important notion in QIT and QT is a quantum correlation,
which is usually termed by entanglement\cite{horodecki09} of given quantum states.
As shown for last two decades it plays a central role in quantum teleportation\cite{teleportation},
superdense coding\cite{superdense}, quantum cloning\cite{clon}, and quantum cryptography\cite{cryptography,cryptography2}. It is also quantum entanglement, 
which makes the quantum computer\footnote{The current status of quantum computer technology was reviewed in Ref.\cite{qcreview}.} outperform the classical one\cite{computer}. Thus, it is very important to understand how to quantify 
and how to characterize the entanglement.

\subsection{entanglement measures}

For bipartite quantum system many entanglement measures were constructed before such as distillable entanglement\cite{benn96}, entanglement 
of formation (EOF)\cite{benn96}, and relative entropy of entanglement (REE)\cite{vedral-97-1,vedral-97-2}.

The distillable entanglement is defined to quantify how many maximally entangled states can be constructed 
from the copies of the given quantum state in the asymptotic region. Thus, in order to compute the distillable 
entanglement we should find the optimal purification (or distillation) protocol. If, for example, the optimal
protocol generates $n$ maximally entangled states from $m$ copies of the quantum state $\rho$, the distillation
entanglement for $\rho$ is given by 
\begin{equation}
\label{distillable}
D (\rho) = \lim_{m \rightarrow \infty} \frac{n}{m}.
\end{equation}
Although the distillable entanglement is well-defined, its analytical calculation is very difficult because 
it is highly non-trivial task to find the optimal purification protocol except very rare cases\cite{distillation}.

REE of a given quantum state $\rho$ is defined as 
\begin{equation}
\label{ree}
E_R (\rho) = \min_{\sigma \in {\cal D}} S(\rho || \sigma),
\end{equation}
where ${\cal D}$ is a set of separable states and $S(\rho || \sigma)$ is a quantum relative entropy;
that is $S(\rho || \sigma) = \mbox{tr} (\rho \ln \rho - \rho \ln \sigma)$. It is known that $E_R (\rho)$ is 
an upper bound of the distillable entanglement. However, for REE it is also highly non-trivial task to 
find the closest separable state $\sigma$ of the given quantum state $\rho$. Still, therefore, we do not know 
how to compute REE analytically even in the two-qubit system except rare cases\cite{ree}.

The EOF for bipartite pure states is defined as a von Neumann entropy of each party, which is derived 
by tracing out other party. For mixed state it is defined via a convex-roof method\cite{benn96,uhlmann99-1};
\begin{equation}
\label{EoF-0}
E_F (\rho) = \min \sum_j p_j E_F (\psi_j),
\end{equation}
where minimum is taken over all possible pure state decompositions, i.e. 
$\rho = \sum_j p_j \ket{\psi_j} \bra {\psi_j}$, with $0 \leq p_j \leq 1$. The decomposition which minimizes
$\sum_j p_j E_F (\psi_j)$ is called the optimal decomposition. For two-qubit system, 
EOF is expressed as\cite{woot-98} 
\begin{equation}
\label{EoF-1}
E_F (C) = h \left( \frac{1 + \sqrt{1 - C^2}}{2} \right),
\end{equation}
where $h(x)$ is a binary entropy function $h(x) = -x \ln x - (1 - x) \ln (1 - x)$ and $C$ is called the concurrence. For two-qubit pure state
$\ket{\psi} = \psi_{ij} \ket{ij}$ with $(i,j=0,1)$, $C$ is given by
\begin{equation}
\label{concurrence-1}
C = |\epsilon_{i_1 i_2} \epsilon_{j_1 j_2} \psi_{i_1 j_1} \psi_{i_2 j_2}| = 2 |\psi_{00} \psi_{11} - \psi_{01} \psi_{10}|,
\end{equation}
where the Einstein convention is understood and $\epsilon_{\mu \nu}$ is an antisymmetric tensor. For two-qubit 
mixed state $\rho$ the concurrence $C(\rho)$ can be computed by $C = \max(\lambda_1 - \lambda_2 - \lambda_3 - \lambda_4, 0)$, where $\{\lambda_1^2, \lambda_2^2, \lambda_3^2, \lambda_4^2\}$ are eigenvalues of 
$\rho (\sigma_y \otimes \sigma_y) \rho^* (\sigma_y \otimes \sigma_y)$ with decreasing order. Thus, one can compute 
the EOF for all two-qubit states in principle. 

\subsection{Classification of Entanglement}

Although quantification of the entanglement is important, it is equally important to classify the entanglement, i.e., to classify the quantum
states into the different type of entanglement. The most popular classification scheme is a classification through a stochastic local operation and 
classical communication (SLOCC)\cite{bennet00}. If $\ket{\psi}$ and $\ket{\phi}$ are in same SLOCC class, this means that $\ket{\psi}$ and $\ket{\phi}$ can be used to implement same task of quantum information process although the probability of success for this task is different. 
Mathematically, if two $n$-party states $\ket{\psi}$ and $\ket{\phi}$ are in the same SLOCC class, they are related to each other by 
$\ket{\psi} = A_1 \otimes A_2 \otimes \cdots \otimes A_n \ket{\phi}$ with $\{A_j\}$ being arbitrary invertible local operators\footnote{For complete proof on the connection between SLOCC and local operations see Appendix A of Ref.\cite{dur00}.}. Moreover, it is more useful to restrict ourselves to SLOCC transformation where all $\{A_j\}$ belong to 
SL($2$, $C$), the group of $2 \times 2$ complex matrices having determinant equal to $1$. In the three-qubit pure-state system it was 
shown\cite{dur00} that there are six different SLOCC classes, fully-separable, three bi-separable, W, and 
Greenberger-Horne-Zeilinger (GHZ) classes. Subsequently, the classification was extended to the
three-qubit mixed-state system\cite{threeM}.

The SLOCC transformation enables us to construct the entanglement measures for the 
multipartite states. As Ref.\cite{verst03} showed, any 
linearly homogeneous positive function of a pure state that is invariant under determinant $1$ SLOCC operations is an entanglement 
monotone. One can show that the concurrence $C$ in Eq. (\ref{concurrence-1}) is such an entanglement monotone as follows. Let $\ket{\psi} = \psi_{ij} \ket{ij}$
with $i,j=0,1$. Then, $\ket{\tilde{\psi}} \equiv (A \otimes B) \ket{\psi} = \tilde{\psi}_{ij} \ket{ij}$, where 
$\tilde{\psi}_{ij} = \psi_{\alpha \beta} A_{i\alpha} B_{j\beta}$. Using $\epsilon_{ij} M_{i \alpha} M_{j \beta} = (\mbox{det} M) \epsilon_{\alpha\beta}$
for arbitrary matrix $M$, it is easy to show $\epsilon_{i_1 i_2} \epsilon_{j_1 j_2} \tilde{\psi}_{i_1 j_1} \tilde{\psi}_{i_2 j_2} = 
(\mbox{det} A) (\mbox{det} B) \epsilon_{i_1 i_2} \epsilon_{j_1 j_2} \psi_{i_1 j_1} \psi_{i_2 j_2}$, which implies that $C$ is invariant under 
determinant $1$ SLOCC operations.

The theorem in Ref.\cite{verst03}, i.e. {\it a linearly homogeneous positive function that remains invariant 
under determinant $1$ SLOCC operation is an entanglement monotone}, can be applied to the three-qubit system. If $\ket{\psi} = \psi_{ijk} \ket{ijk}$, the invariant monotone
is 
\begin{equation}
\label{three-tangle}
\tau_3 = \bigg|2 \epsilon_{i_1 i_2} \epsilon_{i_3 i_4} \epsilon_{j_1 j_2} \epsilon_{j_3 j_4} \epsilon_{k_1 k_3} \epsilon_{k_2 k_4}
          \psi_{i_1 j_1 k_1} \psi_{i_2 j_2 k_2} \psi_{i_3 j_3 k_3} \psi_{i_4 j_4 k_4} \bigg|^{1/2}.
\end{equation}
This is exactly the same with a square root of the residual entanglement\footnote{In this paper we will call $\tau_3$ three-tangle
and $\tau_3^2$ residual entanglement.} introduced in Ref.\cite{ckw}. The three-tangle (\ref{three-tangle}) has following properties.
If $\ket{\psi}$ is a fully-separable or a partially-separable state, its three-tangle completely vanishes. Thus, $\tau_3$ measures  
the true three-way entanglement. It also gives $\tau_3 (\mbox{GHZ}_3) = 1$ and $\tau_3(\mbox{W}_3) = 0$ to the 
three-way entangled states, where
\begin{equation}
\label{ghz3-w3}
\ket{\mbox{GHZ}_3} \frac{1}{\sqrt{2}} (\ket{000} + \ket{111})                 \hspace{2.0cm}
\ket{\mbox{W}_3} = \frac{1}{\sqrt{3}} ( \ket{001} + \ket{010} + \ket{100}).
\end{equation}
For mixed state quantification of the entanglement is usually defined via a convex-roof method\cite{benn96,uhlmann99-1}. Although
the concurrence for an arbitrary two-qubit mixed state can be, in principle, computed following the procedure introduced in Ref.\cite{woot-98}, 
still we do not know how to compute the three-tangle (or residual entanglement) for an arbitrary three-qubit mixed state. However, the
residual entanglement for several special mixtures were computed in Ref.\cite{tangle}. More recently, the three-tangle for all 
GHZ-symmetric states\cite{elts12-1} was computed analytically\cite{siewert12-1}. 

It is also possible to construct the SLOCC-invariant monotones in the higher-qubit systems. In the higher-qubit systems, however, 
there are many independent monotones, because the number of independent SLOCC-invariant monotones is equal to the degrees of freedom
of pure quantum state minus the degrees of freedom induced by the determinant $1$ SLOCC operations. For example, there are 
$2(2^n - 1) - 6 n$ independent monotones in $n$-qubit system. Thus, in four-qubit system there are six invariant monotones. Among them, it was shown in Ref.\cite{four-way} by making use of the antilinearity\cite{uhlmann99-1} that there are following three independent monotones which measure the true four-way entanglement:
\begin{eqnarray}
\label{four-measure}
& &{\cal F}^{(4)}_1 = (\sigma_{\mu} \sigma_{\nu} \sigma_2 \sigma_2) \bullet (\sigma^{\mu} \sigma_2 \sigma_{\lambda} \sigma_2) \bullet
                      (\sigma_2 \sigma^{\nu} \sigma^{\lambda} \sigma_2)           \nonumber  \\
& &{\cal F}^{(4)}_2 = (\sigma_{\mu} \sigma_{\nu} \sigma_2 \sigma_2) \bullet (\sigma^{\mu} \sigma_2 \sigma_{\lambda} \sigma_2) \bullet (\sigma_2 \sigma^{\nu} \sigma_2 \sigma_{\tau}) \bullet (\sigma_2 \sigma_2 \sigma^{\lambda} \sigma^{\tau})                                                                 \\    \nonumber
& &{\cal F}^{(4)}_3 = \frac{1}{2} (\sigma_{\mu} \sigma_{\nu} \sigma_2 \sigma_2) \bullet (\sigma^{\mu} \sigma^{\nu} \sigma_2 \sigma_2) \bullet (\sigma_{\rho} \sigma_2 \sigma_{\tau} \sigma_2) \bullet
(\sigma^{\rho} \sigma_2 \sigma^{\tau} \sigma_2) \bullet (\sigma_{\kappa} \sigma_2 \sigma_2 \sigma_{\lambda})
\bullet (\sigma^{\kappa} \sigma_2 \sigma_2 \sigma^{\lambda}),
\end{eqnarray}
where $\sigma_1 = \openone_2$, $\sigma_1 = \sigma_x$, $\sigma_2 = \sigma_y$, $\sigma_3 = \sigma_z$, and
the Einstein convention is introduced with a metric $g^{\mu \nu} = \mbox{diag} \{-1, 1, 0, 1\}$. 
Furthermore, 
it was shown in Ref.\cite{oster06-1} that there are following three maximally entangled states in 
four-qubit system:
\begin{eqnarray}
\label{four-maximal}
& &\ket{\Phi_1} = \frac{1}{\sqrt{2}} \big(\ket{0000} + \ket{1111} \big)       \nonumber  \\
& &\ket{\Phi_2} = \frac{1}{\sqrt{6}} \left(\sqrt{2} \ket{1111} + \ket{1000} + \ket{0100} + \ket{0010} + \ket{0001}
                                      \right)                                     \\    \nonumber
& &\ket{\Phi_3} = \frac{1}{2} \big(\ket{1111} + \ket{1100} + \ket{0010} + \ket{0001} \big).
\end{eqnarray}

\begin{center}
\begin{tabular}{c|ccc} \hline \hline
& $\hspace{.2cm}{\cal F}^{(4)}_1 \hspace{.2cm}$ &  $\hspace{.2cm} {\cal F}^{(4)}_2 \hspace{.2cm}$  &  
$\hspace{.2cm} {\cal F}^{(4)}_3 \hspace{.2cm}$  \\  \hline 
$\ket{\Phi_1} \hspace{.2cm}$ & $1$ & $1$ & $\frac{1}{2}$     \\   
$\ket{\Phi_2} \hspace{.2cm}$ & $\frac{8}{9}$ & $0$ & $0$      \\  
$\ket{\Phi_3} \hspace{.2cm}$ & $0$ & $0$ & $1$                 \\  
$\ket{\mbox{W}_4} \hspace{.2cm}$  & $0$  &  $0$  &  $0$                \\   \hline  \hline
\end{tabular}

\vspace{0.1cm}
Table I:${\cal F}^{(4)}_1$, ${\cal F}^{(4)}_2$, and ${\cal F}^{(4)}_3$ of the maximally entangled and 
$\mbox{W}_4$ states.
\end{center}

The measures ${\cal F}^{(4)}_1$, ${\cal F}^{(4)}_2$, and ${\cal F}^{(4)}_3$ of $\ket{\Phi_1}$, $\ket{\Phi_2}$, 
$\ket{\Phi_3}$, and 
\begin{equation}
\label{w-4}
\ket{\mbox{W}_4} = \frac{1}{2} \big( \ket{0111} + \ket{1011} + \ket{1101} + \ket{1110} \big)
\end{equation}
are summarized in Table I. As Table I shows, $\ket{\Phi_1}$ is detected by all measures while 
$\ket{\Phi_2}$ (or $\ket{\Phi_3}$) is detected by only ${\cal F}^{(4)}_1$ (or ${\cal F}^{(4)}_3$). As three-qubit
system, $\ket{\mbox{W}_4}$ is not detected by all measures. 

\subsection{Physical Motivations}
As states earlier, W and GHZ classes represent the true $3$-way entanglement in three-qubit system.
However, the three-tangle $\tau_3$ and residual entanglement $\tau_3^2$ cannot detect the entanglement 
of W class, but yield a maximal value to GHZ class. Then, it is natural to ask how much entanglement is 
detected by $\tau_3$ and $\tau_3^2$ for the rank-$2$ mixture $\rho (p) = p \ket{\mbox{GHZ}_3} \bra{\mbox{GHZ}_3}
+ (1 - p) \ket{\mbox{W}_3} \bra{\mbox{W}_3}$. This was explored in the first reference of Ref.\cite{tangle}, 
whose residual entanglement is 
\begin{eqnarray}
\label{summary1}
\tau_3^2 (\rho(p)) = \left\{ \begin{array}{ll}
0 & \hspace{.5cm}   \mbox{for $0 \leq p \leq p_0$}  \\
g_I (p) & \hspace{.5cm}    \mbox{for $p_0 \leq p \leq p_1$}   \\
g_{II} (p) & \hspace{.5cm}    \mbox{for $p_1 \leq p \leq 1$}
                            \end{array}               \right.
\end{eqnarray}
where
\begin{eqnarray}
\label{summary2}
& &g_I (p) = p^2 - \frac{8 \sqrt{6}}{9} \sqrt{p (1-p)^3}  \hspace{1.0cm}
g_{II} (p) = 1 - (1 - p) \left( \frac{3}{2} + \frac{1}{18} \sqrt{465} \right)
                                                                 \\   \nonumber
& &p_0 = \frac{4 \sqrt[3]{2}}{3 + 4 \sqrt[3]{2}} \sim 0.6269  \hspace{1.0cm}
p_1 = \frac{1}{2} + \frac{3}{310} \sqrt{465} \sim 0.7087.
\end{eqnarray}
Thus, one can say that the the influence of W class is dominant at $0 \leq p \leq p_0$ while influence of GHZ class is dominant at $p_1 \leq p \leq 1$. In the intermediate region $p_0 \leq p \leq p_1$ two classes seem to compete
with each other. For three-tangle similar method can be applied and the result is 
\begin{eqnarray}
\label{summary3}
\tau_3 (\rho_p) = \left\{ \begin{array}{ll}
0 & \hspace{.5cm}   \mbox{for $0 \leq p \leq p_0$}  \\
\frac{p - p_0}{1 - p_0}  &  \hspace{.5cm} \mbox{for $p_0 \leq p \leq 1$.}
                           \end{array}                         \right.
\end{eqnarray}
The expression of $\tau_3$ is much simplier than that of $\tau_3^2$. It is mainly due to the fact that 
$\tau_3$ is a linear invariant under the SLOCC transformation. 

\begin{center}
\begin{tabular}{cc} \hline \hline
SLOCC classification  &  representative states    \\   \hline
$G_{abcd}$ &  $\ket{\Phi_1}$, $\ket{\Phi_2}$, $\ket{\Phi_3}$   \\  
$L_{ab_3}$ & $\ket{\mbox{W}_4}$     \\   
$L_{abc_2}$ & $\ket{0000}$      \\  
$L_{a_2 b_2}$ & $\ket{0110} + \ket{0011}$                 \\  
$L_{a_20_{3\oplus\bar{1}}}$  &  $\ket{0011} + \ket{0101} + \ket{0110}$                \\  
$L_{0_{3\oplus\bar{1}}0_{3\oplus\bar{1}}}$  &  $\ket{0000} + \ket{0111}$               \\
$L_{7 \oplus \bar{1}}$  &  $\ket{0000} + \ket{1011} + \ket{1101} + \ket{1110}$          \\
$L_{a_4}$  &  $\ket{0001} + \ket{0110} + \ket{1000}$                                     \\
$L_{5 \oplus \bar{3}}$  & $\ket{0000} + \ket{0101} + \ket{1000} + \ket{1110}$            \\   \hline  \hline
\end{tabular}

\vspace{0.1cm}
Table II: SLOCC classification of four-qubit system.
\end{center}

One can ask same question to four-qubit system by choosing a rank-$2$ state. 
However, situation is much more complicated because 
the four-qubit system has nine different SLOCC classes\cite{fourP-1}. The nine classes and their 
representative states are summarized in Table II. Thus, there are too many combination to choose the rank-$2$
states. Motivated by the three-qubit case we choose two classes $G_{abcd}$ and $L_{ab_3}$ and construct
the rank-$2$ state  $\rho_j = p \ket{\Phi_j} \bra{\Phi_j} + (1 - p) \ket{\mbox{W}_4} \bra{\mbox{W}_4}$ ($j=1,2,3$). The purpose of this paper is to compute ${\cal F}^{(4)}_j$ and ${\cal G}^{(4)}_j$ ($j=1,2,3$), where ${\cal G}^{(4)}_j$ is a linear entanglement monotone defined as 
\begin{equation}
\label{second-degree}
{\cal G}^{(4)}_1 = \left( {\cal F}^{(4)}_1 \right)^{1/3}           \hspace{1.0cm}
{\cal G}^{(4)}_2 = \left( {\cal F}^{(4)}_2 \right)^{1/4}           \hspace{1.0cm}
{\cal G}^{(4)}_3 = \left( {\cal F}^{(4)}_3 \right)^{1/6}.          
\end{equation}
We will show that as in the three-qubit case the influence of $L_{ab_3}$ class is strong at 
$0 \leq p \leq p_0$, where $p_0$ is dependent on $\ket{\Phi_j}$ and is larger than the corresponding 
$3$-qubit value $0.6269$ for most cases. Of course, with increasing $p$ the influence of $G_{abcd}$ becomes stronger gradually.

The paper is organized as follows. In sections II we derive the entanglement of $\rho_1$, $\rho_2$, and $\rho_3$ analytically. We also derive the optimal decompositions explicitly for each range in $p$.
To check the correctness of our results we use the criterion discussed in Ref.\cite{oster07}, i.e. 
{\it entanglement should be a convex hull of the minimum of the characteristic curves}. In section III we discuss the possible applications of our results. In the same section a brief conclusion is given.

\section{Entanglement of $\rho_j \hspace{.2cm} (j = 1, 2, 3)$}

In this section we will compute the entanglement of $\rho_j = p \ket{\Phi_j} \bra{\Phi_j} + (1 - p) \ket{\mbox{W}_4} \bra{\mbox{W}_4}$. Before explicit calculation it is convenient to discuss the general 
method of our calculation briefly. First, we define a pure state
\begin{equation}
\label{revise1}
\ket{Z_j (p, \varphi)} = \sqrt{p} \ket{\Phi_j} - e^{i \varphi} \sqrt{1 - p} \ket{\mbox{W}_4}.
\end{equation}
Since $\ket{Z_j (p, \varphi)}$ is a pure state, one can compute ${\cal F}^{(4)}_i \hspace{.2cm} (i=1,2,3)$
of it by making use of Eq. (\ref{four-measure}). As shown below, ${\cal F}^{(4)}_i$ has a nontrivial zero
at $p = p_0$ for most cases. This is due to the fact that ${\cal F}^{(4)}_i$ cannot detect the entanglement 
of $\ket{\mbox{W}_4}$. Exploiting this fact we construct the optimal decomposition, which yields 
${\cal F}^{(4)}_i (\rho_j) = 0$ at $0 \leq p \leq p_0$. At $p_0 \leq p \leq 1$ region we conjecture the
optimal decomposition and corresponding ${\cal F}^{(4)}_i (\rho_j)$ by making use of the continuity of 
entanglement with respect to $p$ and convex condition. Same procedure can be applied to the computation of 
${\cal G}^{(4)}_i (\rho_j) \hspace{.2cm} (i,j=1,2,3)$. Finally, we adopt a numerical method, which gurantees
the correctness of our guess. 

\subsection{Case $\rho_1$}

In this subsection we will compute the entanglement of $\rho_1 = p \ket{\Phi_1} \bra{\Phi_1} + (1 - p) \ket{\mbox{W}_4} \bra{\mbox{W}_4}$. 
One can show  
\begin{eqnarray}
\label{rho1-F123}
& &{\cal F}^{(4)}_1 \left[Z_1 (p, \varphi)\right] = p |p^2 - 3 (1 - p)^2 e^{4 i \varphi}|    \nonumber \\
& &{\cal F}^{(4)}_2 \left[Z_1 (p, \varphi)\right] = p^2 |p^2 - 4 (1 - p)^2 e^{4 i \varphi}|    \\   \nonumber
& &{\cal F}^{(4)}_3 \left[Z_1 (p, \varphi)\right] = \frac{p^6}{2},
\end{eqnarray}
where $\ket{Z_1 (p, \varphi)}$ is defined in Eq. (\ref{revise1}) with $j=1$.

\subsubsection{${\cal F}^{(4)}_1 \left( \rho_1 \right)$ and ${\cal G}^{(4)}_1 \left( \rho_1 \right)$}
From Eq. (\ref{rho1-F123}) one can show that ${\cal F}^{(4)}_1 \left[Z_1 (p, \varphi)\right]$ has a nontrivial zero
($\varphi = 0$)
\begin{equation}
\label{nont-1}
p_0 = \frac{\sqrt{3}}{\sqrt{3} + 1} \approx 0.634.
\end{equation}
The existence of finite $p_0$ guarantees that ${\cal F}^{(4)}_1 \left( \rho_1 \right)$ should vanish at 
$0 \leq p \leq p_0$. At $p = p_0$ this fact can be verified because we have the optimal decomposition
\begin{eqnarray}
\label{rho1-cr}
& &\rho_1(p_0) = \frac{1}{4} \bigg[ \ket{Z_1 \left(p_0, 0\right)} \bra{Z_1 \left(p_0, 0\right)} + 
\ket{Z_1 \left(p_0, \frac{\pi}{2} \right)} \bra{Z_1 \left(p_0, \frac{\pi}{2} \right)}              \\   \nonumber
& & \hspace{2.5cm}  + 
\ket{Z_1 \left(p_0, \pi \right)} \bra{Z_1 \left(p_0, \pi \right)} + 
\ket{Z_1 \left(p_0, \frac{3 \pi}{2} \right)} \bra{Z_1 \left(p_0, \frac{3 \pi}{2} \right)} \bigg].
\end{eqnarray}
At the region $0 \leq p < p_0$, ${\cal F}^{(4)}_1 \left( \rho_1 \right)$ should vanish too because one can find 
the following optimal decomposition
\begin{equation}
\label{F1-small}
\rho_1 (p) = \frac{p}{p_0} \rho_1 (p_0) + \left(1 - \frac{p}{p_0} \right) \ket{\mbox{W}_4} \bra{\mbox{W}_4}.
\end{equation}
Combining these facts, one can conclude that ${\cal F}^{(4)}_1 \left( \rho_1 \right) = 0$ at $0 \leq p \leq p_0$.

Next, we consider the $p_0 \leq p \leq 1$ region. Eq. (\ref{rho1-cr}) at $p = p_0$ strongly suggests that the 
optimal decomposition at this region is 
\begin{eqnarray}
\label{F1-optimal}
& &\rho_1(p) = \frac{1}{4} \bigg[ \ket{Z_1 \left(p, 0\right)} \bra{Z_1 \left(p, 0\right)} + 
\ket{Z_1 \left(p, \frac{\pi}{2} \right)} \bra{Z_1 \left(p, \frac{\pi}{2} \right)}              \\   \nonumber
& & \hspace{2.5cm}  + 
\ket{Z_1 \left(p, \pi \right)} \bra{Z_1 \left(p, \pi \right)} + 
\ket{Z_1 \left(p, \frac{3 \pi}{2} \right)} \bra{Z_1 \left(p, \frac{3 \pi}{2} \right)} \bigg].
\end{eqnarray}
If Eq. (\ref{F1-optimal}) is a correct optimal decomposition in this region, ${\cal F}^{(4)}_1 \left( \rho_1 \right)$ 
reduces to 
\begin{equation}
\label{F1-large}
{\cal F}^{(4)}_1 \left( \rho_1 \right) = p (6p - 2 p^2 - 3).
\end{equation}
Since the right-hand side of Eq. (\ref{F1-large}) is convex, our conjecture (Eq. (\ref{F1-optimal})) seems to be right. 
In conclusion, we can write
\begin{equation}
\label{F1-all}
{\cal F}^{(4)}_1 \left( \rho_1 \right) = \theta(p - p_0) p (6p - 2 p^2 - 3),
\end{equation}
where $\theta(x)$ is a step function defined as 
\begin{eqnarray}
\label{step}
\theta (x) = \left\{    \begin{array}{cc}
                       1 & \hspace{.5cm} x \geq 0           \\
                       0 & \hspace{.5cm} x < 0.
                         \end{array}            \right.
\end{eqnarray}

However, if our choice Eq. (\ref{F1-optimal}) is incorrect, Eq. (\ref{F1-all}) is merely an upper bound of 
${\cal F}^{(4)}_1 \left( \rho_1 \right)$. Thus, we need to prove that Eq. (\ref{F1-all}) is really optimal value. To prove this we should examine, in principle, all possible decompositions of $\rho_1$, 
i.e. $\rho_1 = \sum_i p_i \ket{\psi_i} \bra{\psi_i}$, and minimize the corresponding value 
$\sum_i p_i {\cal F}^{(4)}_1 (\psi_i)$. However, it is impossible because $\rho_1$ has infinite number of
decomposition.

In order to escape this difficulty to some extent one may rely on Caratheodory's theorem for convex
hull\cite{Caratheodory}, which states that 
for four-qubit rank-$2$ states five vector decomposition is sufficient to minimize ${\cal F}^{(4)}_1 (\rho_1)$. 
Thus, we need to 
investigate decompositions with $2$, $3$, $4$, or $5$ vectors. This method was used in the first reference of 
Ref.\cite{tangle} to minimize the residual entanglement of three-qubit rank-$2$ mixture. Still, however, it is 
difficult, at least for us, to parametrize all decompositions with first few vectors.

In this paper, therefore, we will 
adopt the alternative numerical method presented in Ref.\cite{oster07}. We plot the $p$-dependence of 
${\cal F}^{(4)}_1 \left[Z_1 (p, \varphi)\right]$ for various $\varphi$ (See solid lines of Fig. 1(a)). These 
curves have been referred as the characteristic curves. As Ref.\cite{oster07} showed, 
${\cal F}^{(4)}_1 \left( \rho_1 \right)$ is a convex hull of the minimum of the characteristic curves. Fig. 1(a)
indicates that Eq.(\ref{F1-all}) (thick dashed line) is really the convex characteristic curve, which implies that 
Eq.(\ref{F1-all}) is really optimal. This method was also used in the third reference of 
Ref.\cite{tangle} to minimize the residual entanglement of three-qubit rank-$3$ mixture.

Now, let us consider ${\cal G}^{(4)}_1 \left( \rho_1 \right)$. It is easy to show that 
${\cal G}^{(4)}_1 \left( \rho_1 \right)$ vanishes at $0 \leq p \leq p_0$ due to the optimal 
decomposition Eq. (\ref{F1-small}). If one chooses Eq. (\ref{F1-optimal}) as an optimal decomposition at 
$p_0 \leq p \leq 1$, the resulting ${\cal G}^{(4)}_1 \left( \rho_1 \right)$ is not convex in the full range. 
Thus, we should adopt a technique introduced in Ref.\cite{tangle}. In this case the optimal decomposition is 
\begin{equation}
\label{rho1-G1-optimal}
\rho_1 (p) = \frac{p - p_0}{1 - p_0} \ket{\Phi_1} \bra{\Phi_1} + \frac{1 - p}{1 - p_0} \rho_1 (p_0),
\end{equation}
which results in ${\cal G}^{(4)}_1 \left( \rho_1 \right) = (p - p_0) / (1 - p_0)$. Combining all these facts, 
one can conclude 
\begin{equation}
\label{G1-all}
{\cal G}^{(4)}_1 \left( \rho_1 \right) = \theta(p - p_0) \frac{p - p_0}{1 - p_0}.
\end{equation}

\begin{figure}[ht!]
\begin{center}
\includegraphics[height=5.4cm]{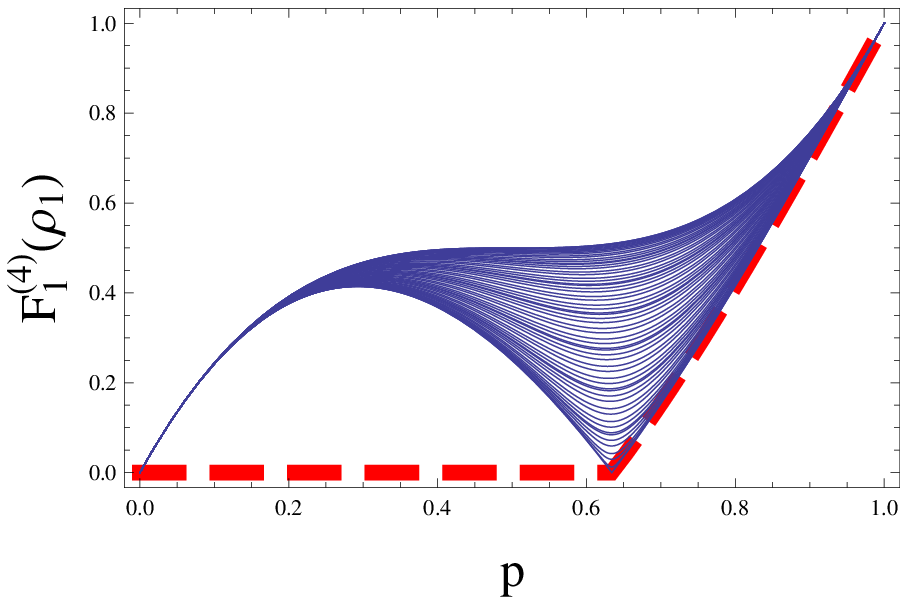}
\includegraphics[height=5.4cm]{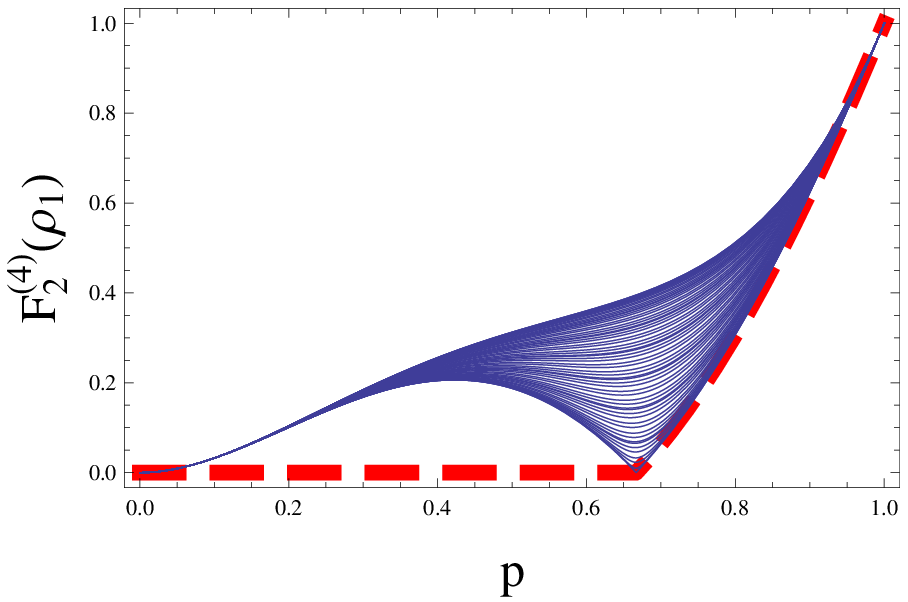}
\includegraphics[height=5.4cm]{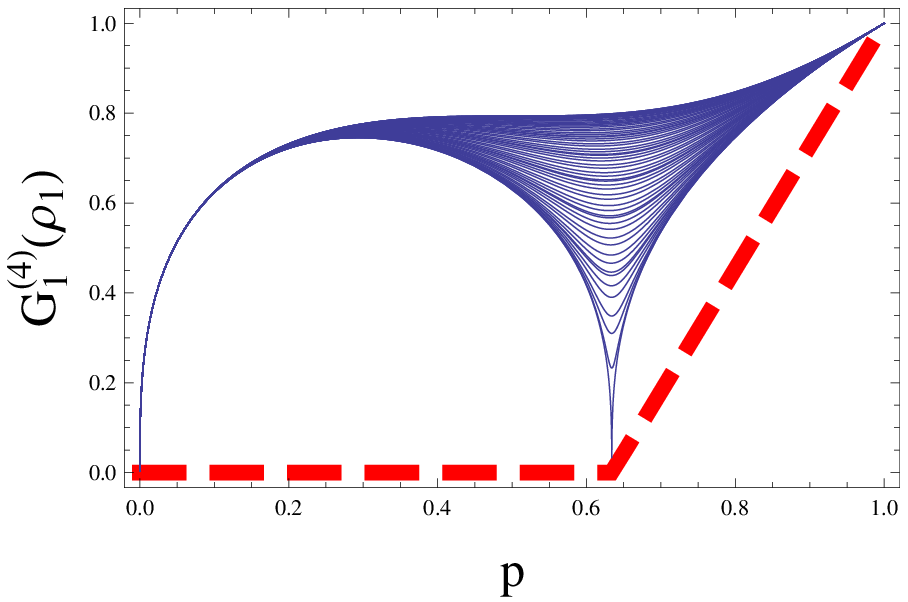}
\caption[fig1]{(Color online) Plot of the $p$ dependence of (a) ${\cal F}^{(4)}_1 \left[Z_1 (p, \varphi)\right]$,
(b) ${\cal F}^{(4)}_2 \left[Z_1 (p, \varphi)\right]$, and (c) ${\cal G}^{(4)}_1 \left[Z_1 (p, \varphi)\right]$ in Eq. (\ref{rho1-F123}). We have chosen $\varphi$ from $0$
to $2\pi$ as an interval $0.1$. The thick dashed lines correspond to ${\cal F}^{(4)}_1 \left( \rho_1 \right)$
in Eq. (\ref{F1-all}), ${\cal F}^{(4)}_2 \left( \rho_1 \right)$ in Eq. (\ref{F2-all}) and 
${\cal G}^{(4)}_1 \left( \rho_1 \right)$ in Eq. (\ref{G1-all}). These figures indicate that 
Eq. (\ref{F1-all}), Eq. (\ref{F2-all}), and Eq. (\ref{G1-all}) are convex hull of the minimum of the characteristic curves.}
\end{center}
\end{figure}

To confirm that Eq. (\ref{G1-all}) is correct, we plot the characteristic curves 
${\cal G}^{(4)}_1 \left[Z_1 (p, \varphi)\right]$ for various $\varphi$ as solid lines and 
Eq. (\ref{G1-all}) as thick dashed line in Fig. 1(c). This figure shows that Eq. (\ref{G1-all}) is convex hull of 
the minimum of the characteristic curves, which strongly supports the validity of Eq. (\ref{G1-all}).

\subsubsection{${\cal F}^{(4)}_2 \left( \rho_1 \right)$ and ${\cal G}^{(4)}_2 \left( \rho_1 \right)$}
From Eq. (\ref{rho1-F123}) one can notice that ${\cal F}^{(4)}_2 \left[Z_1 (p, \varphi)\right]$ has a nontrivial zero
($\varphi = 0$)
\begin{equation}
\label{nont-2}
p_0 = \frac{2}{3} \approx 0.667.
\end{equation}
Thus, Eq. (\ref{F1-small}) and Eq. (\ref{F1-optimal}) with $p_0 = 2/3$ can be the optimal decompositions 
for ${\cal F}^{(4)}_2 \left( \rho_1 \right)$ at $0 \leq p \leq p_0$ and $p_0 \leq p \leq 1$, respectively.
Then, the resulting ${\cal F}^{(4)}_2 \left( \rho_1 \right)$ becomes
\begin{equation}
\label{F2-all}
{\cal F}^{(4)}_2 \left( \rho_1 \right) = \theta(p - p_0) p^2 [p^2 - 4 (1 - p)^2].
\end{equation}

In order to confirm that our result (\ref{F2-all}) is correct, we plot the characteristic curves for various 
$\varphi$ (solid lines) and Eq. (\ref{F2-all}) (thick dashed line) in Fig. 1 (b). As Fig. 1(b) exhibits, 
our result (\ref{F2-all}) is convex hull of the minimum of the characteristic curves, which strongly supports that
Eq. (\ref{F2-all}) is really optimal one.

Similarly, ${\cal G}^{(4)}_2 \left( \rho_1 \right)$ becomes Eq. (\ref{G1-all}) with changing only $p_0$ to $2/3$.
The corresponding optimal decompositions are Eq. (\ref{F1-small}) at $0 \leq p \leq p_0$ and 
Eq. (\ref{rho1-G1-optimal}) at $p_0 \leq p \leq 1$, respectively. Of course, we have to change $p_0$ to $2/3$.

\subsubsection{${\cal F}^{(4)}_3 \left( \rho_1 \right)$ and ${\cal G}^{(4)}_3 \left( \rho_1 \right)$}
Eq. (\ref{rho1-F123}) shows that {${\cal F}^{(4)}_3 \left[ Z_1 (p, \varphi) \right]$ doe not have nontrivial zero. In addition,
it is independent of the phase angle $\varphi$. This fact may indicate that there are infinite number of optimal 
decompositions for ${\cal F}^{(4)}_3 \left( \rho_1 \right)$. The simplest one is
\begin{equation}
\label{F3-optimal}
\rho_1 (p) = \frac{1}{2} \ket{Z_1 (p,0)} \bra{Z_1 (p,0)} +  \frac{1}{2} \ket{Z_1 (p,\pi)} \bra{Z_1 (p,\pi)},
\end{equation}
which gives ${\cal F}^{(4)}_3 \left( \rho_1 \right) = p^6 / 2$. If one chooses Eq. (\ref{F3-optimal}) as an 
optimal decomposition for ${\cal G}^{(4)}_3 \left( \rho_1 \right)$, it generates 
${\cal G}^{(4)}_3 \left( \rho_1 \right) = p /2^{1/6}$. Since it is not concave, we do not need to adopt a technique
to make ${\cal G}^{(4)}_3 \left( \rho_1 \right)$ convex as we did previously. We summarize our results in Table III.

\begin{center}
\begin{tabular}{cccc} \hline  \hline
$j$ & $\hspace{.2cm}{\cal F}^{(4)}_j \hspace{.2cm}$ & $\hspace{.2cm}{\cal G}^{(4)}_j \hspace{.2cm}$ & $p_0$   \\ \hline 
$j=1$ \hspace{.2cm} &  \hspace{.1cm} $p (6 p - 2 p^2 - 3) \theta(p - p_0)$ \hspace{.1cm} & \hspace{.1cm} $\frac{p - p_0}{1 - p_0} \theta(p - p_0)$ \hspace{.1cm} & \hspace{.2cm} $\frac{\sqrt{3}}{\sqrt{3} + 1} \approx 0.634$   
                                                                                                   \\  
$j=2$  \hspace{.2cm} & \hspace{.1cm} $p^2 [p^2 - 4 (1 - p)^2] \theta(p - p_0)$ \hspace{.1cm} & \hspace{.1cm} $\frac{p - p_0}{1 - p_0} \theta(p - p_0)$ \hspace{.1cm} & $\frac{2}{3} \approx 0.667$                     \\   
$j=3$  \hspace{.2cm} & \hspace{.1cm} $\frac{p^6}{2}$ \hspace{.1cm} & \hspace{.1cm} $\frac{p}{2^{1/6}}$ \hspace{.1cm} &                                                             \\   \hline  \hline
\end{tabular}

\vspace{0.1cm}
Table III:Summary of $\hspace{.2cm}{\cal F}^{(4)}_j$ and $\hspace{.2cm}{\cal G}^{(4)}_j$ for $\rho_1$
\end{center}

\subsection{Case $\rho_2$}
In this subsection we would like to quantify the entanglement of $\rho_2$. Above all, we should say that Table I 
implies 
\begin{equation}
\label{rho2-1}
{\cal F}^{(4)}_2 \left( \rho_2 \right) = {\cal G}^{(4)}_2 \left( \rho_2 \right) = 
{\cal F}^{(4)}_3 \left( \rho_2 \right) = {\cal G}^{(4)}_3 \left( \rho_2 \right) = 0,
\end{equation}
because $\rho_2 = p \ket{\Phi_2} \bra{\Phi_2} + (1 - p) \ket{\mbox{W}_4} \bra{\mbox{W}_4}$ itself is an optimal 
decomposition for those entanglement measures. This fact is due to the fact that 
${\cal F}^{(4)}_2$ and ${\cal F}^{(4)}_3$ cannot detect both $\ket{\Phi_2}$ and $\ket{\mbox{W}_4}$. 

\begin{figure}[ht!]
\begin{center}
\includegraphics[height=5.4cm]{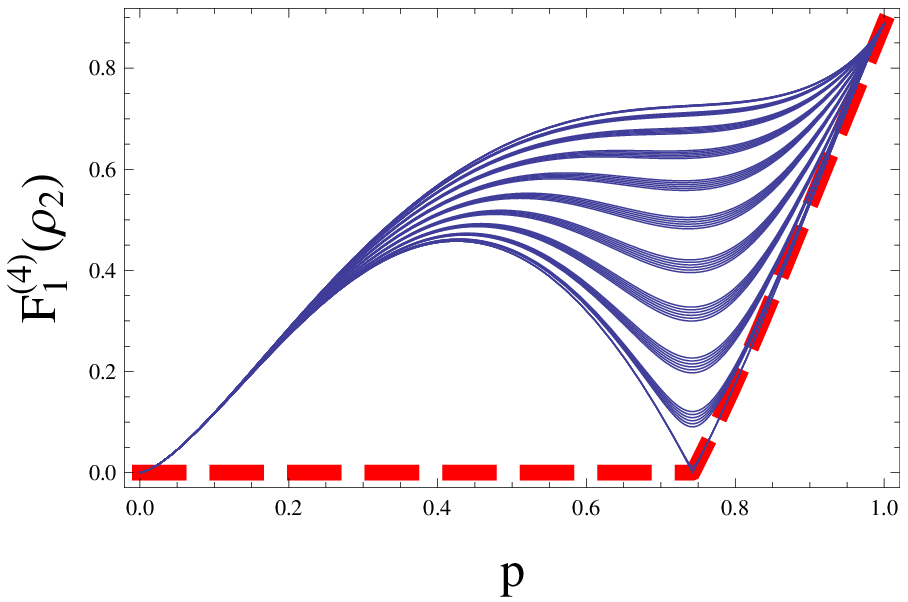}
\includegraphics[height=5.4cm]{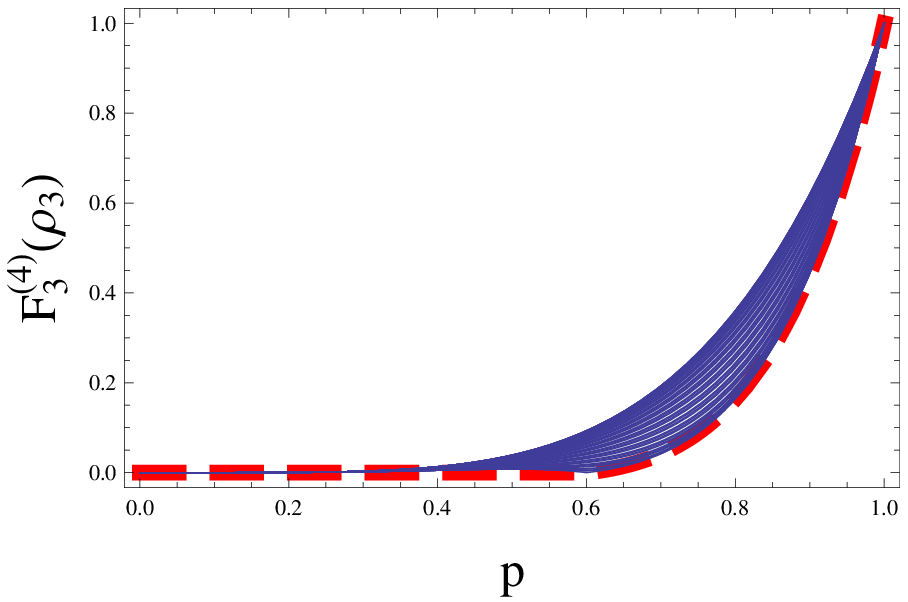}
\caption[fig2]{(Color online) Plot of the $p$ dependence of (a) ${\cal F}^{(4)}_1 \left[Z_2 (p, \varphi)\right]$ in 
Eq. (\ref{rho2-F1})
and (b) ${\cal F}^{(4)}_3 \left[Z_3 (p, \varphi)\right]$ in Eq. (\ref{rho3-F3}). We have chosen $\varphi$ from $0$
to $2\pi$ as an interval $0.1$. The thick dashed lines correspond to ${\cal F}^{(4)}_1 \left( \rho_2 \right)$
in Eq. (\ref{rho2-F1-all}) and ${\cal F}^{(4)}_3 \left( \rho_3 \right)$ in Eq. (\ref{rho3-F3-all}). These figures indicate that 
Eq. (\ref{rho2-F1-all}) and Eq. (\ref{rho3-F3-all}) are convex hull of the minimum of the characteristic curves.}
\end{center}
\end{figure}

Let us now compute ${\cal F}^{(4)}_1 \left( \rho_2 \right)$ and ${\cal G}^{(4)}_1 \left( \rho_2 \right)$. 
It is straightforward to show 
\begin{equation}
\label{rho2-F1}
{\cal F}^{(4)}_1 \left[ Z_2 (p, \varphi) \right] = \frac{8}{9} p^{3/2} |p^{3/2} - 2 \sqrt{6} (1-p)^{3/2} e^{3 i \varphi}|,
\end{equation}
where $\ket{Z_2 (p, \varphi)}$ is given in Eq. (\ref{revise1}).
We notice that ${\cal F}^{(4)}_1 \left[ Z_2 (p, \varphi) \right]$ has a nontrivial zero ($\varphi = 0$)
\begin{equation}
\label{rho2-nont-1}
p_0 = \frac{(2 \sqrt{6})^{2/3}}{1 + (2 \sqrt{6})^{2/3}} \approx 0.743.
\end{equation}
Thus, ${\cal F}^{(4)}_1 \left( \rho_2 \right)$ vanishes at $0 \leq p \leq p_0$ because one can fine the 
optimal decomposition
\begin{equation}
\label{rho2-F1-small}
\rho_2 (p) = \frac{p}{p_0} \rho_2 (p_0) + \left( 1 - \frac{p}{p_0} \right) \ket{\mbox{W}_4} \bra{\mbox{W}_4},
\end{equation}
where 
\begin{equation}
\label{rho2-cr}
\rho_2 (p_0) = \frac{1}{3} \left[ \ket{Z_2 \left(p_0, 0 \right)} \bra{Z_2 \left(p_0, 0 \right)} 
+ \ket{Z_2 \left(p_0, \frac{2\pi}{3} \right)} \bra{Z_2 \left(p_0, \frac{2\pi}{3} \right)}
+ \ket{Z_2 \left(p_0, \frac{4\pi}{3} \right)} \bra{Z_2 \left(p_0, \frac{4\pi}{3} \right)} \right].
\end{equation}

As the previous cases, we adopt, as a trial, the optimal decomposition at $p_0 \leq p \leq 1$ as 
\begin{equation}
\label{rho2-optimal-1}
\rho_2 (p) = \frac{1}{3} \left[ \ket{Z_2 \left(p, 0 \right)} \bra{Z_2 \left(p, 0 \right)} 
+ \ket{Z_2 \left(p, \frac{2\pi}{3} \right)} \bra{Z_2 \left(p, \frac{2\pi}{3} \right)}
+ \ket{Z_2 \left(p, \frac{4\pi}{3} \right)} \bra{Z_2 \left(p, \frac{4\pi}{3} \right)} \right].
\end{equation}
Then ${\cal F}^{(4)}_1 \left( \rho_2 \right)$ becomes $g_I (p)$, where
\begin{equation}
\label{rho2-F1-middle}
g_I (p) =  \frac{8}{9} p^{3/2} \left[p^{3/2} - 2 \sqrt{6} (1-p)^{3/2} \right].
\end{equation}
However, $g_I (p)$ is not convex at the region $p \geq p_{*} \approx 0.9196$. Thus, we should adopt the 
technique previously used again to make $g_I (p)$ convex at the large-$p$ region.

Now, we define $p_1$ such as $p_0 \leq p_1 \leq p_*$. The parameter $p_1$ will be determined later. At the region 
$p_1 \leq p \leq 1$ we adopt the optimal decomposition for ${\cal F}^{(4)}_1 \left( \rho_2 \right)$ as a 
following form:
\begin{eqnarray}
\label{rho2-optimal-2}
\rho_2 (p) = \frac{p - p_1}{1 - p_1} \ket{\Phi_2} \bra{\Phi_2}
+ \frac{1 - p}{1 - p_1} \rho_2 (p_1),
\end{eqnarray}
where 
\begin{equation}
\label{rho2-cr-2}
\rho_2 (p_1) = \frac{1}{3} \left[ \ket{Z_2 \left(p_1, 0 \right)} \bra{Z_2 \left(p_1, 0 \right)} 
+ \ket{Z_2 \left(p_1, \frac{2\pi}{3} \right)} \bra{Z_2 \left(p_1, \frac{2\pi}{3} \right)}
+ \ket{Z_2 \left(p_1, \frac{4\pi}{3} \right)} \bra{Z_2 \left(p_1, \frac{4\pi}{3} \right)} \right].
\end{equation}
Eq. (\ref{rho2-optimal-2}) leads ${\cal F}^{(4)}_1 \left( \rho_2 \right)$ to $g_{II} (p)$ at the 
large-$p$ region, where
\begin{equation}
\label{rho2-F1-large}
g_{II} (p) = \frac{8}{9} \left[ \frac{p - p_1}{1 - p_1} + \frac{1 - p}{1 - p_1}
\left\{ p_1^3 - 2 \sqrt{6} p_1^{3/2} (1 - p_1)^{3/2} \right\} \right].
\end{equation}
As expected $g_{II} (p)$ is convex at $p_1 \leq p \leq 1$.
The parameter $p_1$ is determined by $\frac{\partial g_{II}}{\partial p_1} = 0$, which yields 
$p_1 \approx 0.861$\footnote{The parameter $p_1$ is obtained by an equation $6 p_1 (4 p_1 - 3)^2 = (1 - p_1) (1 + 2 p_1)^2$.}.
Thus, ${\cal F}^{(4)}_1 \left( \rho_2 \right)$ can be summarized as 
\begin{eqnarray}
\label{rho2-F1-all}
{\cal F}^{(4)}_1 \left( \rho_2 \right) = \left\{    \begin{array}{cc}
                                                 0  &  \hspace{.5cm} 0 \leq p \leq p_0          \\
                                                 g_I (p)  & \hspace{.5cm} p_0 \leq p \leq p_1         \\
                                                 g_{II} (p)  & \hspace{.5cm} p_1 \leq p \leq 1.
                                                     \end{array}     \right.
\end{eqnarray}

In order to confirm again that Eq. (\ref{rho2-F1-all}) is correct, we plot the $p$-dependence of the 
characteristic curves (solid lines) in Fig. 2(a) for various $\varphi$. Our result (\ref{rho2-F1-all}) is plotted as a thick dashed line. This figure shows that our result (\ref{rho2-F1-all}) is a convex characteristic curve, 
which strongly supports that our result (\ref{rho2-F1-all}) is correct.

Now, let us compute ${\cal G}^{(4)}_1 \left( \rho_2 \right)$. At $0 \leq p \leq p_0$, 
${\cal G}^{(4)}_1 \left( \rho_2 \right)$ should be zero due to Eq. (\ref{rho2-F1-small}). If we adopt
Eq. (\ref{rho2-optimal-1}) as an optimal decomposition ${\cal G}^{(4)}_1 \left( \rho_2 \right) = g_I^{1/3} (p)$ is obtained. However, it is not convex in the full range. Therefore, we have to choose  
\begin{equation}
\label{rho2-G1-optimal}
\rho_2 (p) = \frac{p - p_0}{1 - p_0} \ket{\Phi_2} \bra{\Phi_2} + \frac{1 - p}{1 - p_0} \rho_2 (p_0)
\end{equation}
as an optimal decomposition, which results in
\begin{equation}
\label{rho2-G1-all}
{\cal G}^{(4)}_1 \left( \rho_2 \right) = \theta(p - p_0) \left(\frac{8}{9}\right)^{1/3} \frac{p - p_0}{1 - p_0}.
\end{equation}

\subsection{Case $\rho_3$}
In this subsection we will compute the entanglement of $\rho_3 = p \ket{\Phi_3} \bra{\Phi_3} + (1 - p) \ket{\mbox{W}_4}
\bra{\mbox{W}_4}$. Since ${\cal F}^{(4)}_1$ and ${\cal F}^{(4)}_2$ cannot detect both $\ket{\Phi_3}$ and $\ket{\mbox{W}_4}$,
it is easy to show 
\begin{equation}
\label{rho3-1}
{\cal F}^{(4)}_1 \left( \rho_3 \right) = {\cal G}^{(4)}_1 \left( \rho_3 \right) = 
{\cal F}^{(4)}_2 \left( \rho_3 \right) = {\cal G}^{(4)}_2 \left( \rho_3 \right) = 0.
\end{equation}

Now, let us compute ${\cal F}^{(4)}_3 \left( \rho_3 \right)$ and ${\cal G}^{(4)}_3 \left( \rho_3 \right)$. 
For $\ket{Z_3 (p,\varphi)}$ in Eq. (\ref{revise1})
it is possible to show that ${\cal F}^{(4)}_3 \left[  Z_3(p, \varphi) \right]$ reduces to 
\begin{equation}
\label{rho3-F3}
{\cal F}^{(4)}_3 \left[ Z_3(p, \varphi) \right] = p^5 \left| p - \frac{3}{2} (1 - p) e^{2 i \varphi} \right|.
\end{equation}
Eq. (\ref{rho3-F3}) implies that ${\cal F}^{(4)}_3 \left[ Z_3(p, \varphi) \right]$ has a nontrivial zero ($\varphi=0$)
\begin{equation}
\label{rho3-nont}
p_0 = \frac{3}{5} = 0.6.
\end{equation}
Thus,  ${\cal F}^{(4)}_3 \left( \rho_3 \right)$ should be zero at the region $0 \leq p \leq p_0$ and its optimal 
decomposition is 
\begin{equation}
\label{rho3-F3-small}
\rho_3 (p) = \frac{p}{p_0} \rho_3 (p_0) + \left( 1 - \frac{p}{p_0} \right) \ket{\mbox{W}_4} \bra{\mbox{W}_4},
\end{equation}
where 
\begin{equation}
\label{rho3-cr}
\rho_3 (p_0) = \frac{1}{2} \left[ \ket{Z_3 (p_0, 0)} \bra{Z_3 (p_0, 0)} + \ket{Z_3 (p_0, \pi)} \bra{Z_3 (p_0, \pi)} \right].
\end{equation}
If we adopt the optimal decomposition at $p_0 \leq p \leq 1$ as a form
\begin{equation}
\label{rho3-optimal-1}
\rho_3 (p) = \frac{1}{2} \left[ \ket{Z_3 (p, 0)} \bra{Z_3 (p, 0)} + \ket{Z_3 (p, \pi)} \bra{Z_3 (p, \pi)} \right],
\end{equation}
the resulting ${\cal F}^{(4)}_3 \left( \rho_3 \right)$ becomes $\frac{5}{2} p^5 \left(p - \frac{3}{5} \right)$. Since this 
is convex, we conclude
\begin{equation}
\label{rho3-F3-all}
{\cal F}^{(4)}_3 \left( \rho_3 \right) = \theta (p - p_0) \frac{5}{2} p^5 \left(p - \frac{3}{5} \right).
\end{equation}
In order to prove that Eq. (\ref{rho3-F3-all}) is correct we plot again the characteristic curves (solid lines) and 
our result  (\ref{rho3-F3-all}) (thick dashed line) in Fig. 2(b), which supports that Eq. (\ref{rho3-F3-all}) is optimal one.

Finally, let us compute ${\cal G}^{(4)}_3 \left( \rho_3 \right)$. If we take Eq. (\ref{rho3-optimal-1}) as an 
optimal decomposition for ${\cal G}^{(4)}_3 \left( \rho_3 \right)$ at $p_0 \leq p \leq 1$, the result is not convex
in the full range of this region. Thus, we should choose
\begin{equation}
\label{rho3-G3-optimal}
\rho_3 (p) = \frac{p - p_0}{1 - p_0} \ket{\Phi_3} \bra{\Phi_3} + \frac{1 - p}{1 - p_0} \rho_3 (p_0)
\end{equation}
as an optimal decomposition, which simply results in the right-hand side of Eq. (\ref{G1-all}) with $p_0 = 3 / 5$. 

\section{Discussion and Conclusions}

We compute the three-kinds of true four-way entanglement measures ${\cal F}^{(4)}_j \hspace{.2cm} (j=1,2,3)$ and
their corresponding linear entanglement monotones ${\cal G}^{(4)}_j \hspace{.2cm} (j=1,2,3)$ analytically
for four-qubit rank-$2$ mixed states $\rho_j = p \ket{\Phi_j} \bra{\Phi_j} + (1 - p) \ket{\mbox{W}_4} \bra{\mbox{W}_4} \hspace{.2cm} (j=1,2,3)$. All optimal decompositions consist of $2$, $3$, $4$, and $5$
vectors. 

Our results can be used to find many different mixed states, which have vanishing entanglement. For example, 
let us consider ${\cal F}^{(4)}_1$ with $p_0$ in Eq. (\ref{rho2-nont-1}). Let us represent, for simplicity, 
$\ket{\Phi_2}$ and $\ket{\mbox{W}_4}$ as 
\begin{eqnarray}
\label{concl-1}
\ket{\Phi_2} = \left( \begin{array}{c} $1$ \\ $0$  \end{array} \right)          \hspace{1.0cm}
\ket{\mbox{W}_4} = \left( \begin{array}{c} $0$ \\ $1$  \end{array} \right).
\end{eqnarray}
Imagine the two-dimensional space spanned by $\ket{\Phi_2}$ and $\ket{\mbox{W}_4}$ represented by a 
Bloch sphere. Then, the states in the Bloch sphere can be expressed as 
$\rho=\frac{1}{2} (\openone + \bm{r} \cdot \bm{\sigma})$, where $|\bm{r}| = 1$ and $|\bm{r}| < 1$ denote
the pure and mixed states, respectively. In this representation the Bloch vectors of $\ket{\Phi_2}$, 
$\ket{\mbox{W}_4}$, and $\ket{Z_2 (p_0, \varphi)}$ are
\begin{eqnarray}
\label{concl-2}
& &\hspace{2.0cm}\bm{r} (\Phi_2) = (0, 0, 1)           \hspace{.5cm}
\bm{r} (\mbox{W}_4) = (0, 0, -1)                           \\   \nonumber
& &\bm{r} (Z_2 (p_0, \varphi)) = (-2 \sqrt{p_0 (1 - p_0)} \cos \varphi, -2 \sqrt{p_0 (1 - p_0)} \sin \varphi, 
                                2 p_0 - 1).
\end{eqnarray}
Thus, any states located in the tetrahedron , whose vertices are $(0, 0, -1)$, 
$(-2 \sqrt{p_0 (1 - p_0)}, 0, 2p_0-1)$, $(\sqrt{p_0 (1 - p_0)}, - \sqrt{3p_0 (1 - p_0)}, 2 p_0 - 1)$, 
and  $(\sqrt{p_0 (1 - p_0)}, \sqrt{3p_0 (1 - p_0)}, 2 p_0 - 1)$ in the Bloch sphere representation, have
vanishing ${\cal F}^{(4)}_1$ and ${\cal G}^{(4)}_1$.

\begin{center}
\begin{tabular}{c||c|c} \hline
& ${\cal C}$ (concurrence)  &  $\tau$ (three-tangle)                 \\   \hline
$\rho_1$ \hspace{.3cm} &  $\frac{1}{2} \left(1 - 2 \sqrt{p} - p \right) \theta (\alpha_1 - p)$ \hspace{.2cm} $\left(\alpha_1 = (\sqrt{2} - 1)^2 \right)$ &  0    \\   \hline
$\rho_2$ \hspace{.3cm} & $ \left( \frac{3 - p}{6} - \frac{\sqrt{2}}{3} \sqrt{p (3-p)} \right) \theta (\alpha_2 - p)$ 
\hspace{.2cm} $\left(\alpha_2 = \frac{1}{3} \right)$ &  ?
                                                                                            \\   \hline
  &  ${\cal C}_{AB} = \frac{1}{2} \left(1 - 2 \sqrt{p} - p \right) \theta (\alpha_1 - p)$  &     \\ 
$\rho_3$ \hspace{.3cm} &  ${\cal C}_{AC} = {\cal C}_{AD} = {\cal C}_{BC} = {\cal C}_{BD}$  &  $\tau_{ACD} = \tau_{BCD} = 0$  \\
  &  $ = \frac{1}{2} \left( 1 - p - \sqrt{p (2 - p)} \right) \theta (\alpha_3 - p)$  
 \hspace{.2cm} $\left(\alpha_3 = \frac{2 - \sqrt{2}}{2} \right)$ &  
                                                                                 $\tau_{ABC} = \tau_{ABD} = ?$   \\
  &  ${\cal C}_{CD} = \frac{1}{2} \left\{ 1 - \sqrt{\frac{p}{2}} \left( \sqrt{1 + \sqrt{p (2 - p)}} + 
                              \sqrt{1 - \sqrt{p (2 - p)}}  \right)   \right\}$  &                    \\   \hline
\end{tabular}

\vspace{0.1cm}
Table IV:Entanglement for sub-states of $\rho_j \hspace{.2cm} (j=1,2,3)$.
\end{center}

One can use our results to discuss the monogamy properties\cite{Eltschka04} of entanglement. 
For this purpose, however, 
we should compute the entanglement for the sub-states of $\rho_j \hspace{.2cm} (j=1,2,3)$. The entanglement
of the sub-states is summarized at Table IV. As this table shows, some three-tangle, at least for us, cannot be
computed analytically. This is because still we do not have a closed formula for computing the three-tangles.

As mentioned above, there are nine SLOCC classes in the four-qubit system. Therefore, many rank-$2$ states 
can be constructed by choosing different classes. If, for example, $G_{abcd}$ and $L_{7 \oplus \bar{1}}$ are
chosen, one can construct the rank-$2$ state
\begin{equation}
\label{pij}
\pi_j = p \ket{\Phi_j} \bra{\Phi_j} + (1 - p) \ket{\xi} \bra{\xi},
\end{equation} 
where $\ket{\xi} = (\ket{0000} + \ket{1011} + \ket{1101} + \ket{1110}) / 2$. Probably, our calculation
procedure enable us to compute the entanglement of $\pi_2$ and $\pi_3$ although we have not checked it
explicitly. For $\pi_1$, however, our procedure does not seem to work because of 
$\bra{\xi} \Phi_1 \rangle \neq 0$. 
Using Table II one can construct many higher rank states. If, for example, $G_{abcd}$, $L_{7 \oplus \bar{1}}$,
and $L_{a_4}$ are chosen, one can construct the rank-$3$ state such as 
\begin{equation}
\label{sigmaj}
\sigma_j = p \ket{\Phi_j} \bra{\Phi_j} + q \ket{\xi} \bra{\xi} + (1 - p - q) \ket{\eta} \bra{\eta},
\end{equation}
where $\ket{\eta} = (\ket{0001} + \ket{0110} + \ket{1000}) / \sqrt{3}$. However, it seems to be highly
difficult to compute the entanglement of higher rank states.

The most remarkable achievement and novelty of this paper is deriving the entanglement of four-qubit mixed states using an analytical approach.
Thus, our result may serve as a quantitative reference for future studies of entanglement in quadripartite
and/or multipartite mixed states.

{\bf Acknowledgement}:
This research was supported by the Basic Science Research Program through the National Research Foundation of Korea(NRF) funded by the Ministry of Education, Science and Technology(2011-0011971).

\end{document}